\documentclass[a4paper,12pt]{article}
\def\letter{0}\def\pr{0}
\pdfoutput=1 
\usepackage{graphicx}
\usepackage{epstopdf}
\usepackage{ifthen}
\usepackage{csquotes}
\usepackage[numbers,sort&compress]{natbib}
\usepackage{tikz}
\usetikzlibrary{arrows.meta} 

\usepackage{amsmath}
\usepackage[psamsfonts]{amssymb}
\usepackage{euscript}
\usepackage{caption}
\usepackage{latexsym}
\usepackage[arrow,matrix,curve]{xy}

\usepackage[hypertexnames=false]{hyperref}
\hypersetup
{
    colorlinks=true,
    linkcolor=blue,
    filecolor=magenta,   
    citecolor=black,   
    urlcolor=cyan,
}

\def\,{\hspace{-.1cm}}
\def\hsp{,\hspace{.7cm}}

\def\fc#1#2 {\frac{n}{q}#1\frac{n}{q}#2}

\renewcommand{\sinh}{\textrm{sinh}}
\renewcommand{\cosh}{\textrm{cosh}}
\renewcommand{\tanh}{\textrm{tanh}}
\newcommand{\sech}{\textrm{sech}}
\newcommand{\csch}{\textrm{csch}}

\renewcommand{\theequation}{\arabic{section}.\arabic{equation}}
\renewcommand{\(}{\begin{equation}}
\renewcommand{\)}{end{equation} \vspace{-.05in}\linebreak}

\newcounter{saveeqn}
\newcounter{savealpheqn}

\newcommand{\alpheqn}{\setcounter{saveeqn}{\value{equation}}%
  \stepcounter{saveeqn}\setcounter{equation}{0}%
  \renewcommand{\theequation}{\mbox{\arabic{section}.\arabic{saveeqn}
\alph{equation}}}
  \renewcommand{\)}{\end{equation}}}
\def\part#1{\frac{\partial}{\partial{#1}}}%
\def\group#1{\refstepcounter{equation}\setcounter{saveeqn}
 {\value{equation}}%
  \label{#1}\setcounter{equation}{0}%
\renewcommand{\theequation}{\mbox{\arabic{section}.\arabic{saveeqn}
\alph{equation}}}
  \renewcommand{\)}{\end{equation}}}
\newcommand{\reseteqn}{\setcounter{equation}{\value{saveeqn}}%
  \renewcommand{\theequation}{\arabic{section}.\arabic{equation}}%
  \renewcommand{\)}{\end{equation}}}

\newcommand{\aalpheqn}{\setcounter{saveeqn}{\value{equation}}%
  \stepcounter{saveeqn}\setcounter{equation}{0}%
  \renewcommand{\theequation}{\mbox{
        \Alph{subsection}.\arabic{saveeqn}\alph{equation}}}
   \renewcommand{\)}{\end{equation}}}
\newcommand{\areseteqn}{\setcounter{equation}{\value{saveeqn}}%
  \renewcommand{\theequation}{\Alph{subsection}.\arabic{equation}}%
  \renewcommand{\)}{\end{equation}}}

\renewcommand{\thefootnote}{\alph{footnote}}
\renewcommand{\(}{\begin{equation}}
\renewcommand{\)}{\end{equation}}
\newcommand{\ba}{\begin{eqnarray}}
\newcommand{\ea}{\end{eqnarray}}
\newcommand{\cbp}{\mathop{\vtop{\ialign{##\crcr
   $\hfil\displaystyle{}\hfil$\crcr\noalign{\kern-13pt\nointerlineskip}
   \BIG{)}\hskip0pt\crcr\noalign{\kern3pt}}}}}
\newcommand{\pa}{\mathop{\vtop{\ialign{##\crcr

$\hfil\displaystyle{\oplus}\hfil$\crcr\noalign{\kern+1pt\nointerlineskip
}
   \hspace{.08in}$^{\alpha=0}$\hskip6pt\crcr\noalign{\kern3pt}}}}}
\renewcommand{\hsp}{,\hspace{.3in}}
\newcommand{\p}{^\prime}

\catcode`\@=11
\def\vereq#1#2{\lower3pt\vbox{\baselineskip1.5pt \lineskip1.5pt
\ialign{$\m@th#1\hfill##\hfil$\crcr#2\crcr\sim\crcr}}}
\catcode`\@=12

\renewcommand{\(}{\begin{equation}}
\renewcommand{\)}{\end{equation}}

\def\k#1{\k_{#1}}

\def\g{\mathfrak g}

\def\gre#1{\textcolor{magenta}{jarah: #1}}
\usepackage[dvipsnames]{xcolor}

\newcommand{\beas}{\begin{eqnarray*}}
\newcommand{\eeas}{\end{eqnarray*}}

\newcommand{\bquo}{\begin{quote}}
\newcommand{\enqu}{\end{quote}}

\def\lim#1{\stackrel{\rm{lim}}{{}_{#1}}}

\def\ok#1{\omega_{k_{#1}}}

\newcommand{\beq}{\begin{equation}}
\newcommand{\eeq}{\end{equation}}
\newcommand{\bea}{\begin{eqnarray}}
\newcommand{\eea}{\end{eqnarray}}

\newif\ifdtup

\catcode`\@=11

\ifthenelse{\equal{\letter}{0}}
{ 

\@addtoreset{equation}{section}
\def\theequation{\arabic{section}.\arabic{equation}}

\def\@normalsize{\@setsize\normalsize{15pt}\xiipt\@xiipt
\abovedisplayskip 14pt plus3pt minus3pt%
\belowdisplayskip \abovedisplayskip
\abovedisplayshortskip \z@ plus3pt%
\belowdisplayshortskip 7pt plus3.5pt minus0pt}

\def\small{\@setsize\small{13.6pt}\xipt\@xipt
\abovedisplayskip 13pt plus3pt minus3pt%
\belowdisplayskip \abovedisplayskip
\abovedisplayshortskip \z@ plus3pt%
\belowdisplayshortskip 7pt plus3.5pt minus0pt
\def\@listi{\parsep 4.5pt plus 2pt minus 1pt
      \itemsep \parsep
      \topsep 9pt plus 3pt minus 3pt}}

\relax

\def\section{\@startsection{section}{1}{\z@}{3.5ex plus 1ex minus  .2ex}{2.3ex plus .2ex}{\large\bf}}

\def\thesection{\arabic{section}}
\def\thesubsection{\arabic{section}.\arabic{subsection}}

\def\appendix{\setcounter{section}{0}
 \def\thesection{Appendix \Alph{section}}
 \def\thesubsection{\Alph{section}.\arabic{subsection}}
 \def\theequation{\Alph{section}.\arabic{equation}}}
\renewcommand{\theequation}{\arabic{section}.\arabic{equation}}
}
{
\renewcommand{\theequation}{\arabic{equation}}
} 

\begin{document}
\def\thefootnote{\fnsymbol{footnote}}
\def\thetitle
{(De-)Exciting the Third P\"oschl-Teller Kink}
\def\autone{Hengyuan Guo}
\def\auttwo{Jarah Evslin}
\def\autthree{Stefano Bolognesi}


\def\affa{ School of Physics and Astronomy, Sun Yat-sen University, Zhuhai 519082, China}
\def\affaa{Department of Physics, ``Enrico Fermi'', University of Pisa; INFN, Sezione di Pisa, \\ Largo Pontecorvo, 3, 56127, Pisa, Italy}
\def\affaaa{Lanzhou Center for Theoretical Physics, \\ Key Laboratory of Theoretical Physics of Gansu Province, \linebreak Key Laboratory of Quantum Theory and Applications of MoE,  Lanzhou University, \\ Lanzhou, Gansu 730000, China}
\def\affb{Institute of Modern Physics, NanChangLu 509, Lanzhou 730000, China}
\def\affc{University of the Chinese Academy of Sciences, YuQuanLu 19A, Beijing 100049, China}



\ifthenelse{\equal{\pr}{1}}
{
\title{\thetitle}
\author{\autone}
\author{\auttwo}
\author{\autthree}
\affiliation {\affa}
\affiliation {\affaa}
\affiliation {\affaaa}
\affiliation {\affb}
\affiliation {\affc}
}

\begin{center}
{\large {\bf \thetitle}}

\bigskip

\large \noindent  
\autone$^{1,2,3}$  
\footnote{guohy57@mail.sysu.edu.cn},  
\auttwo$^{4,5}$ 
\footnote{jarah@impcas.ac.cn}
and  
\autthree$^{2}$
\footnote{stefano.bolognesi@unipi.it}

\vskip.7cm  
{\small 1) \affa\\[5pt]  
2) \affaa\\[5pt]
3) \affaaa\\[5pt]
4) \affb\\[5pt]
5) \affc}
\end{center}

\begin{abstract}
\noindent
There is a series of scalar models possessing reflectionless kinks whose linear perturbations are described by a P\"oschl-Teller potential at integer level $\sigma$.  The cases $\sigma=1$ and $2$ are the well-known Sine-Gordon and $\phi^4$ double-well models.  The $\sigma=3$ kink has received relatively little attention because it exhibits a $\phi^{8/3}$ potential, whose third derivative  diverges in the vacuum.  In old-fashioned perturbation theory this yields a cubic interaction that diverges far from a kink.  We nonetheless use this interaction to calculate the amplitudes and probabilities for incoming radiation to excite or de-excite one of the kink's two shape modes.  As each shape mode is localized about the kink, the leading order amplitudes are nonetheless finite.  This suggests that the $\sigma=3$ model is not pathological, but rather its mesons are quantum field theoretic extensions of Znojil's bound states.

\end{abstract}

%
\setcounter{footnote}{0}
\renewcommand{\thefootnote}{\arabic{footnote}}

\ifthenelse{\equal{\pr}{1}}
{
\maketitle
}{}

\section{Introduction}
\subsection{Motivation}
The study of quantum solitons burst to life when Ref.~\cite{dhn2} found the one-loop mass correction to the kink in the $\phi^4$ double-well model.  A flurry of papers \cite{sg1,sg2,sg3,sg4,sg5} generalized this result to the soliton of the Sine-Gordon model.

By now, much is known about the quantum properties of the kinks in both models.  In the case of the Sine-Gordon model this is to a significant extent due to its integrability.  But in both cases a key to progress has been that the linearized perturbations of the kink are described by the exactly solvable P\"oschl-Teller (PT) potential, respectively at levels $\sigma=2$ and $\sigma=1$.  While kinks in many other models have been extensively studied classically, their quantum lifts are largely unexplored beyond a calculation of their one-loop masses.  In fact many such kinks, such as that of the $\phi^6$ model, do not lift to any Hamiltonian eigenstate because quantum corrections cause the vacuum energies to differ on their two sides \cite{wstabile}.

Needless to say, learning about quantum kinks from only two examples is rather limiting.  For example, the Sine-Gordon soliton has no bound shape modes, and the $\phi^4$ kink only has one.  On the other hand, kinks with multiple shape modes have a rich phenomenology already classically \cite{cabtesi,wob23} but little is known about their quantum field theory lifts. 

It was realized immediately in Ref.~\cite{sg2} that these two examples are part of an infinite series of models whose kinks have linearized perturbations solving PT potentials whose levels are any natural number $\sigma$.  However, it was noted in that reference that, when the level $\sigma>2$, the potential ``contains unphysical singularities."  Ref.~\cite{tf} went further, writing, ``The singularities in their higher derivatives
can be expected to cause difficulties in calculations
which incorporate these derivatives; for example: high-order
perturbation theories of kink response to external forces."  The purpose of the present paper is to test this expectation, albeit at tree-level.

Despite these unphysical singularities, the one-loop mass corrections to the kinks were computed in Ref.~\cite{boya} and were found to be finite.  In fact, the kinks were shown to be linearly stable in Ref.~\cite{Alonso-Izquierdo:2012csn}. 

In the present note, we provide more evidence that these kinks are not pathological.  The potential is known not to be singular, it suffers merely from a lack of analyticity at the minima corresponding to the vacua, where the third derivative of the potential diverges.  As a result, care must be taken when using the usual power-series expansion based perturbation theory.

More precisely, expanding about the minimum of the potential, the leading interaction term is of the form $\phi^{8/3}$ in the case $\sigma=3$ and $\phi^{5/2}$ in the case $\sigma=4$.  These functions certainly do not diverge at finite $\phi$.  However, the three-point function in the vacuum, derived by taking the third derivative of the potential, does diverge.  At any finite distance from the kink, $\phi$ is not in its vacuum state and so even the three-point function is finite, but one must check to see whether any given observable diverges.  The present paper considers one observable, an inelastic scattering amplitude of a kink with a quantum of radiation, and finds that while this unbounded potential is indeed integrated, the integral turns out to be finite and so the resulting amplitude is finite.  This lends evidence to the conjecture that these models are well behaved, thus greatly enlarging the set of kink models with exactly solvable perturbations to infinity.

Indeed, in quantum mechanics, such rational power-law potentials are quite familiar.  In Ref.~\cite{znojil} the bound states were found, and we suspect that the analogue of the vacuum sector Fock space consists of lifts of these bound states to quantum field theory.

Our project is also motivated by various applications in higher-dimensional and complex backgrounds such as \cite{Zhong:2021jhep,Zhong:2023pel,Ringe:2024ktt}. Rational power-law potentials, such as those in the PT kink models studied here, are increasingly used in models from inflationary cosmology~\cite{frac1,frac2,frac3}, where they yield small but measurable tensor perturbations \cite{ibe} and primordial black holes~\cite{pbh24}, to $T\overline{T}$ deformed field theories \cite{smirnov16,torino16} where square roots in the flowed spectrum of an apparently Hermitian Hamiltonian have led to deep conceptual puzzles.  Recently, the phenomenology of kinks in models with various nonanalytic potentials has been studied \cite{na0,na1,na2,na3} finding rather rich phenomenology, such as phase transitions to regimes of oscillon production \cite{na4}. Furthermore, the presence of multiple shape modes provides a laboratory to extend the meson-baryon scattering program initiated by Uehara and collaborators \cite{Uehara:1985zx}, where the excitation spectrum of the nucleon is derived from the fluctuation modes of the Skyrmion. 

\subsection{Methods}


Perhaps the most powerful and certainly the most popular formalism for treating quantum solitons is the collective coordinate method of Refs.~\cite{cc1,cc2,sg2}. 
It promotes the collective coordinates to operators.   However, their commutation relations are quite complicated, and so one projects the other operators onto their orthogonal compliment, which corresponds to a field redefinition which leads to an infinite number of terms in the Hamiltonian.  As a result, with some notable exceptions \cite{andy1,itoff,vac24,andy2}, in its half-century history it has yielded relatively few novel results.  It also faces numerous potential pitfalls, such as the fact that collective coordinates are generally multivalued~\cite{multi}.

Fortunately, for problems involving a single soliton and a finite number of radiation quanta, which we will call mesons, there is a much simpler method \cite{mekink,me2loop} called Linearized Soliton Perturbation Theory (LSPT).  This method has already been successfully applied to exhaustively calculate the amplitudes of the leading elastic \cite{Evslin:2024jhep} and inelastic \cite{mestokes, Evslin:2023vdw} processes in kink-meson scattering. 

Kink-meson scattering is interesting in its own right.  Early works \cite{uehara91} used kink-meson scattering to learn about baryon-meson scattering phenomenology, as kinks are toy model solitons and baryons become solitons when the number of colors is large.  Kink-radiation interactions have been shown to exhibit interesting phenomenology such as negative radiation pressure \cite{negpress08}.  It was believed that such interactions are responsible for the chaotic resonance windows of Ref.~\cite{Campbell:1983zu} however recent evidence \cite{noshape} suggests a more interesting story \cite{story}.



This formalism has also been applied to higher-dimensional kinks \cite{Guo:2025stokes}.  In this setting, it has long been suspected \cite{0812.1929} that interactions of radiation with the internal excitations of solitons are responsible for the absence of small loops in cosmic string simulations.  Very recent simulations~\cite{2405.06030} have confirmed the key role played by shape mode-radiation interactions.

This paper is organized as follows. In Section \ref{sec:PT} we derive the potential of a model whose kink's perturbations are described by the PT Schrodinger problem, reproducing the results of Ref.~\cite{tf}.  
In Sec.~\ref{sec:scattering} we calculate the Stokes and anti-Stokes  scattering amplitudes and probabilities for the $\sigma=3$ model. In Sec.~\ref{sec:fit} we expand the potential of this model near the minima and find a $\phi^{8/3}$ leading interaction. In Sec.~\ref{sec:discussion}, we discuss present and future results.

\section{Classical Models Yielding P\"oschl-Teller Kinks}
\label{sec:PT}

\subsection{General Models}

We consider a (1+1)-dimensional model of a classical scalar field $\phi(x,t)$ and its conjugate momentum $\pi(x,t)$, described by the Hamiltonian density
\bea
\mathcal{H}=\frac{\pi^2+(\nabla \phi)^2}{2}+V[\phi]. \label{hc}
\eea 
The classical equation of motion is 
\beq
\partial_{\mu}\partial^{\mu}\phi+\frac{\partial V[\phi]}{\partial \phi}=0.
\eeq
We assume that the potential $V[\phi]$ has degenerate trivial vacua $v_{i}$ 
\beq
\frac{\partial V[\phi]}{\partial{\phi}}\Big|_{\phi=v_i}=0\hsp v_{i+1}>v_i \hsp i=1,2,\cdots\ .
\eeq
A kink is a non-trivial static soliton solution  $\phi(x,t)=f(x)$  that satisfies
\beq
\nabla^2 f(x)=\frac{\partial V}{\partial\phi}\Big|_{\phi=f(x)} \hsp f(x) = \begin{cases}
 v_i&\text{if } x\rightarrow{ -\infty}\\
 v_j&\text{if } x\rightarrow{ \infty}   \nonumber
  \end{cases}
 \eeq
We will be interested in BPS kinks, which are minimal energy configurations that also satisfy the first order BPS equation
\beq
\frac{\partial  f(x)}{\partial x}=\sqrt{2V[\phi(x)]}.
\label{BPSeq}
\eeq
Such a solution of the first order equation always exists for one real scalar field and adjacent vacua $j=i+1$.   The classical kink mass is 
\beq
    Q_0
    =\int _{-\infty}^{\infty} dx\left(\frac{1}{2}(\nabla f)^2+V[f]\right)=2\int _{-\infty}^{\infty} dx V[f]. 
\eeq

Now consider a small, periodic perturbation
\beq
\phi(x,t)=f(x)+\g(x)e^{-i\omega t}.
\eeq
The classical equation of motion yields the Sturm-Liouville equation for the perturbation
\beq
\left(-\nabla^2+\frac{\partial^2 V[\phi]}{\partial \phi^2}\Big|_{\phi=f}\right)\g(x)=\omega^2\g(x)+O(\g^2).
\label{solfluc}
\eeq
We will be interested in infinitesimal perturbations, and so we drop the $O(\g^2)$ corrections.

\subsection{P\"oschl-Teller Potentials}

Consider the case in which this Sturm-Liouville equation takes the PT form, up to a constant shift $K$
\beq
(-\nabla^2+U(x))\g(x)=\omega^2\g(x) \hsp U(x)=-\frac{a^2\sigma(\sigma+1)}{\cosh^2(ax)}+K. \label{pt}
\eeq
Here $a$ and $\sigma$ are positive real numbers.

This can be solved using the standard Hamiltonian Factorization method, reviewed in~\ref{app}.  The ground state is
\beq
\g_{B}(x)= c_0 \sech^\sigma(ax)\hsp \omega_0^2=-a^2\sigma^2+K \label{g0}
\eeq
for some normalization constant $c_0$.  We want this to correspond to the zero-mode fluctuation of our kink and so we demand that $\omega_0=0$, implying
\beq
K=a^2\sigma^2.
\eeq
Now the PT potential is labeled by two parameters: the dimensionless $\sigma$ and also the dimensionful scale $a$.  The meson mass is
\beq
m=\sqrt{K}=a\sigma.
\eeq

Note that $\sigma$ can be any positive real number~\cite{Alonso-Izquierdo:2012csn}. When $\sigma$ is an integer, there are $\sigma-1$ discrete modes, provided in \ref{app}, and the last one has eigenvalue $\omega_{\sigma-1}^2=a^2\sigma^2-a^2 $.



\subsection{Soliton with general PT potential}

Identifying the PT Schrodinger equation (\ref{pt}) with the Sturm-Liouville equation (\ref{solfluc}) for the linearized perturbations of a soliton, we conclude that
\beq
\frac{\partial^2V}{\partial \phi}\Big|_f=U(x)=-\frac{m^2(\sigma+1)}{\sigma\cosh^2\left(\frac{mx}{\sigma}\right)}+m^2.
\eeq

The soliton solution $f(x)$ is determined by the fact that $\g_B(x)$ is its translation zero mode.  Normalizing $\g_B(x)$ so that it square integrates to unity by fixing
\beq
c_0=\left[\int_{-\infty}^{+\infty}\sech^{2\sigma}\left(\frac{mx}{\sigma}\right)dx)\right]^{-1/2}
   =  \left(\left(\frac{\sigma y}{m} {}_2F_1\left(\frac{1}{2},1-\sigma,\frac{3}{2},y^2\right)\right|_{-1}^1 \right)^{-1/2}
   =\frac{\sqrt{m\Gamma[\sigma+ \frac{1}{2}]}}{\pi^{1/4}\sqrt{\sigma\Gamma[\sigma]}} 
\eeq
the fact that $\g_B$ is a translation mode implies that
\beq
\g_B(x)=\pm \frac{\partial_x f(x)}{\sqrt{Q_0}}.
\eeq
Ref.~\cite{sg2} noted that that this may be integrated to find $f(x)$, as was done in Ref.~\cite{tf}.  Fixing the sign and constant of integration, one finds the kink profile
\beq
  f(x)=\sqrt{Q_0}\int dx \g_B(x)=\frac{\sqrt{Q_0}\sigma c_0}{m} {}_2F_1\left(\frac{1}{2}, 1-\frac{\sigma}{2}, \frac{3}{2},\tanh^2\left(\frac{mx}{\sigma}\right)\right)\tanh\left(\frac{mx}{\sigma}\right).\label{fx}
\eeq
The BPS equation (\ref{BPSeq})
then provides the potential energy density along the kink
\beq
\begin{aligned}
    V[f(x)]=&\frac{Q_0\g^2_B(x)}{2}=\frac{Q_0}{2}\frac{m\Gamma[\sigma+1/2]}{\pi^{1/2}\sigma\Gamma[\sigma]}\sech^{2\sigma}\left(\frac{mx}{\sigma}\right).\label{Uf}  
\end{aligned}
\eeq

\section{Stokes and Anti-Stokes Scattering}\label{sec:scattering}

\subsection{General Formulas}

We will now restrict our attention to the $\sigma=3$ model, whose kink has two shape modes.  We consider the Schrodinger picture quantum field theory whose Hamiltonian density is that of Eq.~(\ref{hc}) but normal-ordered with the usual Schrodinger picture plane-wave normal ordering prescription.  The unbound modes describe radiation whose quanta will be referred to as mesons.

We will consider two inelastic scattering processes.  The first, Stokes scattering \cite{stokes52,raman28,landsberg28}, is the process in which a meson with initial momentum $k_0$ strikes a ground state kink from the left, exciting one of the two shape modes and then either continuing to the right or else rebounding to the left
\beq
{\rm{kink}}+{\rm{meson}}\rightarrow {\rm{kink}}^*+{\rm{meson}}.
\eeq
The second, anti-Stokes scattering, consists of a meson with initial momentum $k_0$ which strikes a kink with an excited shape mode, de-exciting it
\beq
{\rm{kink}}^*+{\rm{meson}}\rightarrow {\rm{kink}}+{\rm{meson}}.
\eeq
The momentum of the outgoing meson is
\beq
k^S_f(k_0)=\sqrt{(\omega_{k_0}-\omega_S)^2-m^2}\hsp k^{aS}_f(k_0)=\sqrt{(\omega_{k_0}+\omega_S)^2-m^2}
\eeq
in the case of Stokes and anti-Stokes scattering respectively.  Note that
\beq
k^{aS}_f(k^S_f(k_0))=k_0.
\eeq

The respective probabilities of Stokes and anti-Stokes scattering on a general reflectionless kink were computed in Ref.~\cite{mestokes} 
\bea
P_{\rm{S}}(k_0)&=&\lambda 
\frac{|V_{S,k_f^S(k_0),-k_0}|^2+|V_{S,-k_f^S(k_0),-k_0}|^2}{8k_0k_f^S(k_0)\omega_{S}} \label{princ}\\
P_{\rm{aS}}(k_0)&=&\lambda
\frac{|V_{S,k_f^{aS}(k_0),-k_0}|^2+|V_{S,-k_f^{aS}(k_0),-k_0}|^2}{8k_0k_f^{aS}(k_0)\omega_{S}} \nonumber
\eea
where the three-point coupling is
\beq
V_{Sk_1k_2}=\int dx V^{'''}[f(x)]\g_{k_1}(x)\g_{k_2}(x)\g_{S}(x)\label{vkkk}
\eeq
and the two $|V|^2$ terms in the numerator correspond to forward and backward scattering.  Note that
\beq
P_{aS}(k_f^S(k_0))=P_S(k_0)\hsp P_{aS}(k_0)=P_S(k_f^{aS}(k_0))
\eeq
where the reversibility arises from the density of states in two-dimensional kinematics.

Here $S$ is whichever shape mode is excited or de-excited.  The notation $\omega$ is again used for the energy, so that $\omega_S$ is the energy of the shape mode while for continuum modes
\beq
\omega_k=\sqrt{m^2+k^2}.
\eeq

\subsection{Finiteness}
The $V^{'''}$ appearing in the three-point coupling (\ref{vkkk}) is precisely the divergent term discussed in Refs.~\cite{sg2,tf}.  It diverges at $|x|\rightarrow\infty$, corresponding to the early and late times in our scattering process.  In fact, in the derivation of Eq.~(\ref{princ}), the initial meson began in a wave packet centered at $x_0\ll 0$, and so it spent a very long time in this divergent region.  Therefore, one may expect \cite{tf} that this divergence would lead to divergences in the probabilities for various processes involving the meson.

\begin{figure}
    \centering
    \includegraphics[width=0.8\linewidth]{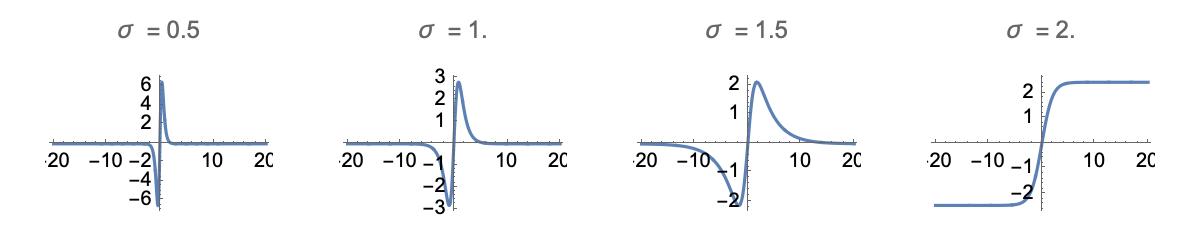} 
    \includegraphics[width=0.8\linewidth]{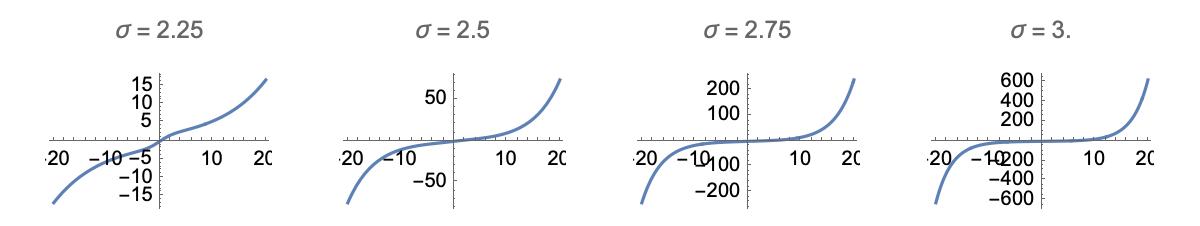}
    \caption{Left to right: $V'''[f(x)]$ at $m=1$ for different values of $\sigma$.}
    \label{fig1}
\end{figure}

More precisely, the third derivative of the potential is 
\bea
  V^{'''}[f(x)]&=&\frac{\partial V^{''}[\phi(x)]}{\partial \phi}|_{f(x)}
  =[\frac{\partial V^{''}[\phi(x)]}{\partial x}\frac{\partial \phi}{\partial_x}]|_{f(x)}\\
  &=&\frac{2\pi^{1/4} (\sigma +1) m^3 \sqrt{\frac{\sigma  \Gamma (\sigma )}{m \Gamma \left(\sigma +\frac{1}{2}\right)}} \tanh \left(\frac{m x}{\sigma }\right) \sech(\frac{m x}{\sigma })^{2-\sigma }}{\sqrt{Q_0}\sigma ^2}.\nonumber
\eea
For the $\sigma > 2$ models, the potential's third derivative $V'''[f(x)]$ exhibits a formal divergence as $x \to \pm \infty$ (as shown in Fig.~\ref{fig1}).  However, the inelastic scattering matrix element,
\begin{equation}
V_{Sk_1k_2} = \int_{-\infty}^{+\infty} dx V'''[f(x)] \g_{k_1}(x) \g_{k_2}(x) \g_S(x),
\end{equation}
remains physically well-defined and convergent. 

The convergence is guaranteed by the localized nature of the shape mode $\g_S(x)$. When $\sigma > 2$, the internal bound state $\g_S(x)$ decays at least exponentially as $e^{-|x|}$. This convergence is sufficiently fast to cancel the divergence of the potential term, rendering the integral finite.

\subsection{The $\sigma=3$ Kink}
In the case $\sigma=3$ of interest here, the kink solution and potential are
\beq
\begin{aligned}
  f(x)=&\frac{3}{8}\sqrt{\frac{5}{m}}\sqrt{Q_0}\bigg(\arcsin(\tanh\left(\frac{mx}{3}\right))+\tanh\left(\frac{mx}{3}\right) \sech\left(\frac{mx}{3}\right)\bigg)\\
  V[f(x)]=&\frac{5}{32} Q_0 m \sech^6\left(\frac{m x}{3}\right)\rightarrow V^{'''}[f(x)]=\frac{32m^{5/2}}{9\sqrt{5Q_0}}\sinh\left(\frac{mx}{3}\right).\label{fV3}
\end{aligned}
\eeq
One sees that $V^{'''}$ diverges as $e^{m|x|/3}$.

\begin{figure}
    \centering
     \includegraphics[width=.8\linewidth]{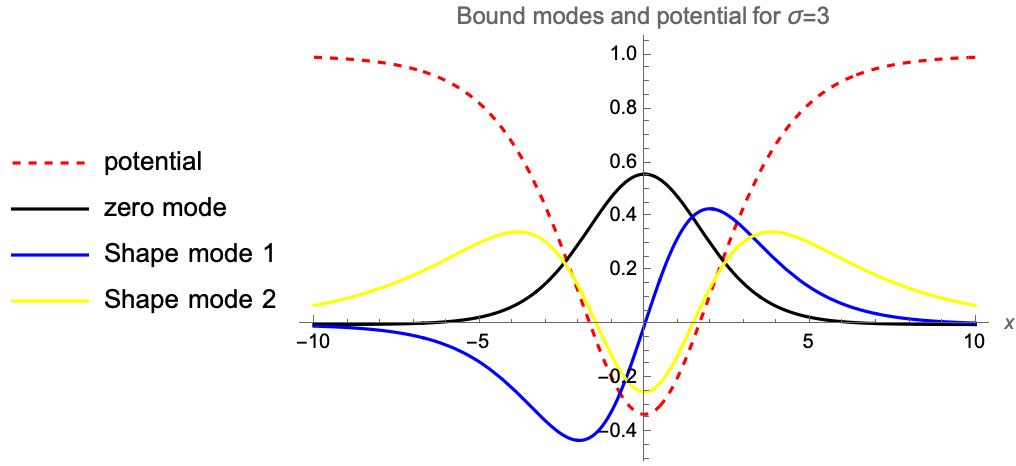} 
        \caption{Two shape modes and one zero mode in the $\sigma=3$ PT potential}
    \label{fig2}
\end{figure}
There are three bound states which are plotted in Fig.~\ref{fig2}.  Together with the continuum states, these are
\bea
\g_B(x)&=&\frac{\sqrt{5m}}{4}\sech^3\left(\frac{mx}{3}\right) \hsp \omega_0=0\\
\g_{S_1}(x)&=&\frac{\sqrt{5m}}{2}\sech^{2}\left(\frac{mx}{3}\right)\tanh\left(\frac{mx}{3}\right)\hsp \omega_{S_1}^2=\frac{5m^2}{9}\nonumber\\ 
\g_{S_2}(x)&=&{\sqrt{m}\left(\sech \left(\frac{m x}{3}\right)-\frac{5}{4}\sech^3\left(\frac{m x}{3}\right)\right) }=\frac{m^2 \left(2 \cosh \left(\frac{2 m x}{3}\right)-3\right) \text{sech}^3\left(\frac{m x}{3}\right)}{4 \sqrt{m^3}}
\hsp \omega_{S_2}^2=\frac{8m^2}{9} \nonumber \\
  \g_k(x)&=&\frac{e^{i k x}}{D_3}  \bigg[ik(9k^2-11 m^2)+2 m(-9k^2+m^2)\tanh\left(\frac{mx}{3}\right) \nonumber\\
  &&+m^2 \sech^2\left(\frac{mx}{3}\right)(-5 m \tanh\left(\frac{mx}{3}\right)+15 i k)\bigg]
\eea
where the normalization factor is
\beq
  D_3[k]=9 i k^3+18 k^2 m-11 i k m^2-2 m^3.
\eeq
Here $\g_B$ is the translation mode while $\g_{S_1}$ and $\g_{S_2}$ are the two shape modes.  Note that $\g_{S_1}$ vanishes at large $|x|$ as $e^{-2m|x|/3}$, more than compensating for the divergence in $V^{'''}$.  However the other shape mode, $\g_{S_2}$, only vanishes as $e^{-m|x|/3}$ and so $|V^{'''}\g_{S_2}|$ tends to a constant.  The phase oscillation in the continuum modes nonetheless makes the three-point function $V_{Skk}$ convergent, as the incoming and outgoing momenta differ.

More precisely, the three-point vertex factors are
\beq
\begin{aligned}
&V_{k_1k_2S_1}=\int dx V^{'''}[f(x)]\g_{k_1}(x)\g_{k_2}(x)\g_{S_1}(x)\\
=&\frac{m\pi\ \sech\left(\frac{3\pi(k_1+k_2))}{2m}\right)}{72\sqrt{Q_0} \left(-9 i k_1^3-18 k_1^2 m+11 i k_1 m^2+2 m^3\right) \left(-9 i k_2^3-18 k_2^2 m+11 i k_2 m^2+2 m^3\right)}\\
\times&\bigg[
-3645 (k_1^8+k_2^8)+2916k_1^2k_2^2(k_1^4+k_2^4)+1458k_1^4k_2^4+1124m^6(k_1^2+k_2^2)+107 m^8\\
&+m^2\left(-3564(k_1^6+k_2^6)+6156k_1^2k_2^2(k_1^2+k_2^2)\right)+m^4(1098 (k_1^4+k_2^4)+5004k_1^2k_2^2)\bigg]
\end{aligned} 
\eeq
and 
\beq
\begin{aligned}
&V_{k_1k_2S_2}=\int dx V^{'''}[f(x)]\g_{k_1}(x)\g_{k_2}(x)\g_{S_2}(x)\\
=
&\frac{im\pi\ \csch\left(\frac{3\pi(k_1+k_2))}{2m}\right)}{144 \sqrt{5Q_0} \left(-9 i k_1^3-18 k_1^2 m+11 i k_1 m^2+2 m^3\right) \left(-9 i k_2^3-18 k_2^2 m+11 i k_2 m^2+2 m^3\right)}\\
&\times\bigg[-2592 m^2 \left(k_1^2+k_2^2\right) \left(5 k_1^4-18 k_1^2 k_2^2+5 k_2^4\right)-3645 \left(5 k_1^4+6 k_1^2 k_2^2+5 k_2^4\right) \left(k_1^2-k_2^2\right)^2\\
&+12544 m^6 \left(k_1^2+k_2^2\right)++144 m^4 \left(109 k_1^4+230 k_1^2 k_2^2+109 k_2^4\right)+2048 m^8\bigg]\\
&-\frac{128 i \pi  m^4 \left(k_1 \left(9 k_1^2-11 m^2\right) \left(m^2-9 k_2^2\right)+k_2 \left(m^2-9 k_1^2\right) \left(9 k_2^2-11 m^2\right)\right) \delta \left(\frac{k_1+k_2}{m}\right)}{27 \sqrt{5Q_0} \left(9 k_1^3-18 i k_1^2 m-11 k_1 m^2+2 i m^3\right) \left(9 k_2^3-18 i k_2^2 m-11 k_2 m^2+2 i m^3\right)}.
\end{aligned}
\eeq
One observes the $\delta$ function divergence in the last term, which would contribute if the incoming and outgoing momenta were equal
.  However the tree-level probability (\ref{princ}) involves only the vertex factors on-shell, where they are not equal.


\subsection{Scattering Probabilities}
In the case of the $i$th shape mode, where $i=$1 or 2, the Stokes and anti-Stokes scattering probabilities in Eq.~(\ref{princ}) are respectively
\bea
P^{(i)}_{\rm{S}}(k_0)|&=&\lambda
\frac{|V_{S_i,\sqrt{(\omega_{k_0}-\omega_{S_i})^2-m^2},-k_0}|^2+|V_{S_i,-\sqrt{(\omega_{k_0}-\omega_{S_i})^2-m^2},-k_0}|^2}{8k_0\omega_{S_i}\sqrt{(\omega_{k_0}-\omega_{S_i})^2-m^2}}\\
P^{(i)}_{\rm{aS}}(k_0)|&=&\lambda
\frac{|V_{S_i,\sqrt{(\omega_{k_0}+\omega_{S_i})^2-m^2},-k_0}|^2+|V_{S_i,-\sqrt{(\omega_{k_0}+\omega_{S_i})^2-m^2},-k_0}|^2}{8k_0\omega_{S_i}\sqrt{(\omega_{k_0}+\omega_{S_i})^2-m^2}}.\nonumber
\eea

\begin{figure}
    \centering
  \includegraphics[width=0.4\linewidth]{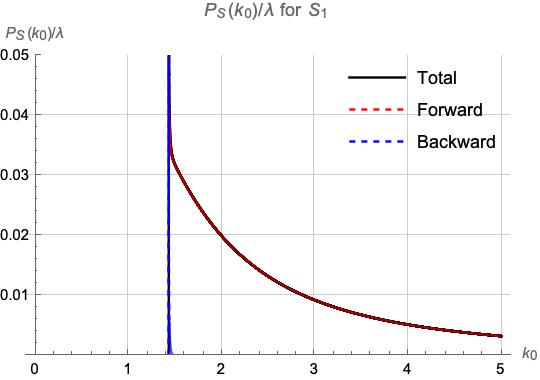}
  \includegraphics[width=0.4\linewidth]{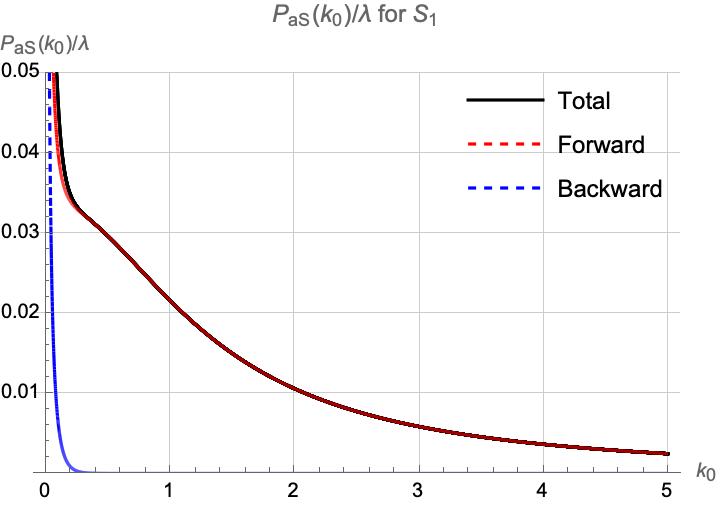}
\includegraphics[width=0.4\linewidth]{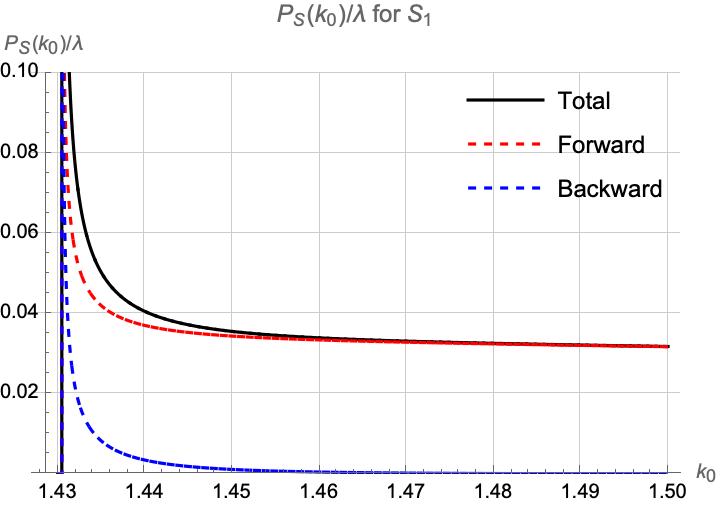} 
\includegraphics[width=0.4\linewidth]{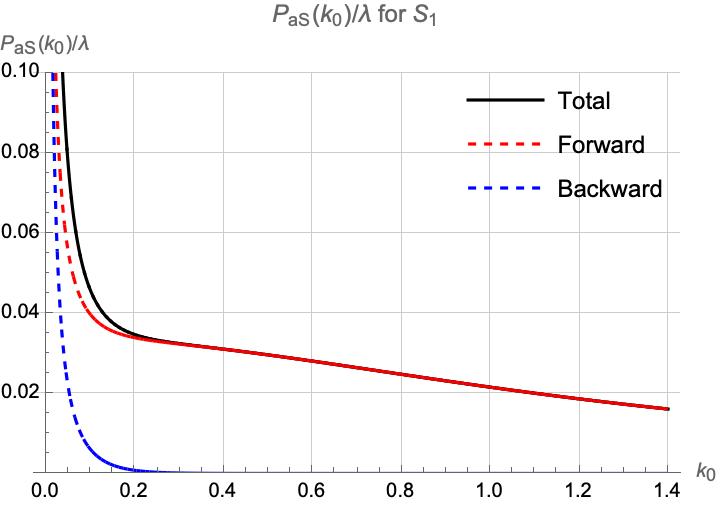} 
    \caption{Stokes (left) and anti-Stokes (right) scattering probabilities as functions of $k_0$ for the first shape mode in the case $\sigma=3$.  The bottom panels are zoomed in on energies just above their respective thresholds.} 
    \label{fig3}
\end{figure}

\begin{figure}
    \centering  
    \includegraphics[width=0.4\linewidth]{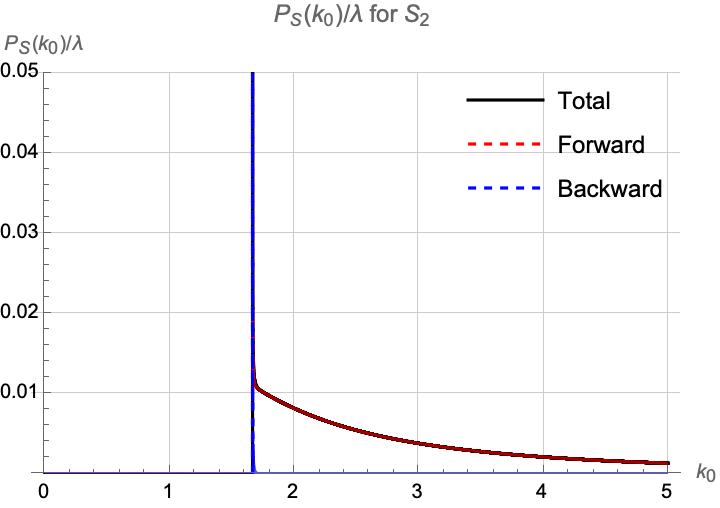}
    \includegraphics[width=0.4\linewidth]{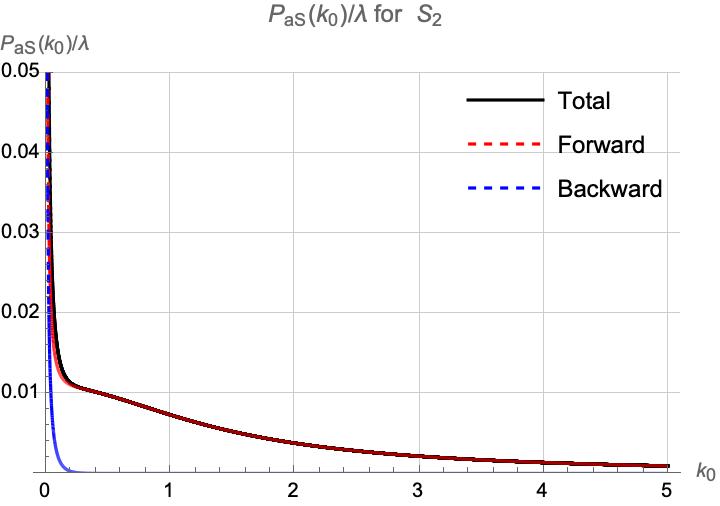} 
    \includegraphics[width=0.4\linewidth]{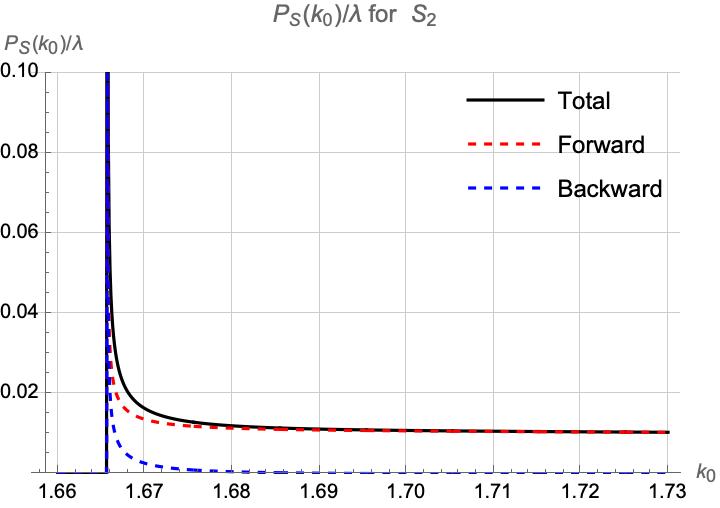}
    \includegraphics[width=0.4\linewidth]{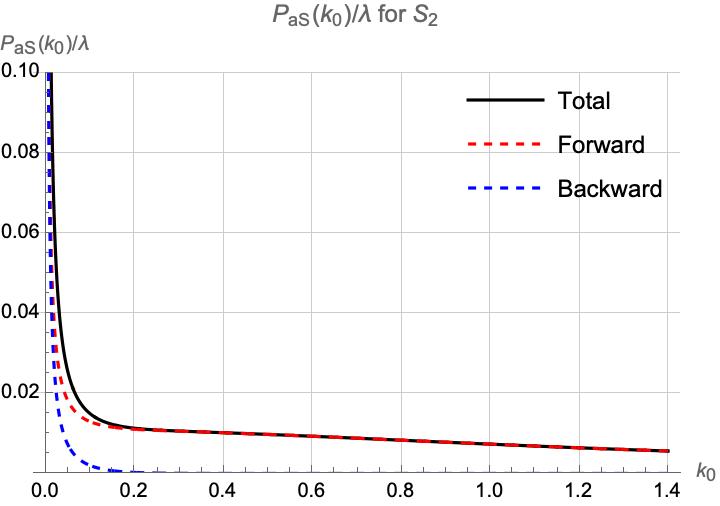}
    \caption{Stokes (left) and anti-Stokes (right) scattering probabilities as functions of $k_0$ for the second shape mode in the case $\sigma=3$.  The bottom panels are zoomed in on energies just above their respective thresholds.}
    \label{fig4}
\end{figure}

These are plotted in Figs.~\ref{fig3} and \ref{fig4} respectively.  One can see that they diverge near the threshold, due to the usual inverse velocity enhancement.  Backward scattering of the outgoing meson is only appreciable near the threshold for each process.


The threshold for Stokes scattering, in the case of each shape mode, is
\beq
\begin{aligned}
 &(\omega_{k0}- \omega_{S_1})^2-m^2\geq 0\rightarrow 
k_{0}\geq \frac{m}{3} \sqrt{5+6 \sqrt{5}}\simeq 1.43048m \\
 &(\omega_{k0}- \omega_{S_2})^2-m^2\geq 0\rightarrow  
 k_{0}\geq \frac{m}{3} \sqrt{2+3 \sqrt{2}}\simeq 1.66569m.
\end{aligned}
\eeq
\par 

\section{Expanding the Potential about its Minimum}\label{sec:fit}
Although in the quantum theory we use a semiclassical expansion which is a limit in which the coupling vanishes, in this section we will return to the classical field theory.  Now that we no longer set $\hbar=1$, the coupling is dimensionful.  Therefore we may, for simplicity, fix the coupling so that
\beq
Q_0=\frac{256m}{45\pi^2}.
\eeq
This is a convenient choice because then the minima of the potential are at $\phi=\pm 1$.  The field and the potential become
\beq
\begin{aligned}
  f(y)=&C_1 \bigg(\sech(y)\tanh(y)+\arcsin(\tanh(y))\bigg)\hsp 
  V[f(x)]=C_2\sech^6(y) \hsp y=\frac{mx}{3}\label{fV3a} 
\end{aligned}
\eeq
where
\beq
f(\pm\infty)=\pm v=\pm 1\hsp C_1=\frac{2}{\pi}\hsp C_2=\frac{8}{9}m^2.
\eeq

\subsection{The Center of the Kink}
Let us first expand about the center of the kink, $y=0$.  Using
\beq
\begin{aligned}
 \arcsin(\tanh(y))=& y-\frac{y^3}{6}+O(y^4)\hsp \sech(y)\tanh(y)= y-\frac{5 y^3}{6}+O\left(y^4\right)\hsp \text{for the kink }\\
 \sech^6(y)=&1-3 y^2+O\left(y^4\right)\hsp \text{for the potential }
\end{aligned}
\eeq
one easily finds 
\beq
\begin{aligned}
  \phi=f= 2C_1y+O(y^3)\hsp 
  V[\phi]= C_2(1-3y^2)+O(y^4)=C_2 - \frac{3C_2}{4C_1^2} \phi^2+O(\phi^4).
\end{aligned} 
\eeq
This is the usual expansion for an interacting, massive scalar.  The potential is symmetric in the middle of the kink, but more generally one also expects a cubic interaction.

\subsection{The Asymptotic Vacua}
Now let us consider $y\rightarrow\infty$ where $\phi$ tends to the $\phi=1$ vacuum.  Let us expand
\beq
\phi=1-\epsilon.
\eeq
Expanding the hyperbolic tangent function
\beq
\begin{aligned}
 \tanh(y)=1-\frac{2e^{-y}}{e^y+e^{-y}}=1-2e^{-2y}+2e^{-4y}+O(e^{-6y})=1-\delta+\frac{\delta^2}{2}+O(\delta^3)\hsp \delta=2e^{-2y}.
\end{aligned}
\eeq
Similarly
\bea
\sech(y)&=&2e^{-y}-2e^{-3y}+O(e^{-5y})=\sqrt{2\delta}-\frac{\delta\sqrt{\delta}}{\sqrt{2}}+O(\delta^2\sqrt{\delta})\\
\tanh(y)\sech(y)&=&\sqrt{2\delta}-\frac{3\delta\sqrt{\delta}}{\sqrt{2}}+O(\delta^2\sqrt{\delta})\nonumber\\
\arcsin(\tanh(y))&=&\arcsin\left(1-\delta+
\frac{\delta^2}{2}+O(\delta^3)\right)=\frac{\pi}{2}-\sqrt{2\delta}+\frac{\delta\sqrt{\delta}}{3\sqrt{2}}+O(\delta^2\sqrt{\delta}).\nonumber
\eea
The kink profile is then
\beq
1-\epsilon=\phi=f=1-\frac{8\sqrt{2}}{3\pi}\delta\sqrt{\delta}+O(\delta^2\sqrt{\delta}).
\eeq
We conclude that the difference between the field value and the vacuum is
\beq
\epsilon=\frac{8\sqrt{2}}{3\pi}\delta\sqrt{\delta}+O(\delta^2\sqrt{\delta})
\eeq
and so
\beq
\delta\propto \epsilon^{2/3}.
\eeq
Now
\beq
V[f(y)]=C_2\sech^6(y)=8c_2\delta^3(1-3\delta)+O(\delta^5)\propto \epsilon^2+O(\epsilon^{8/3}).
\eeq
This is our main result, the difference $\epsilon$ between $\phi$ and vacuum value is subjected to a potential with a usual mass term plus a $\epsilon^{8/3}$ term. 

In the corresponding quantum mechanics, in which one supplements a harmonic oscillator $\omega^2x^2$ term with a $\lambda x^{8/3}$ term, this does not affect the leading order oscillator states.  However, it does affect the states at subleading orders in perturbation theory.  In quantum field theory we expect that it will take them out of the Fock space generated by acting the creation and annihilation operators on the perturbative vacuum, as these fractional powers cannot be written as polynomials in the creation and annihilation operators.  Perhaps they play a role similar to twist operators in orbifold models, leading to the existence of other sectors.  It would be interesting to understand what physical effects result.

In summary, our potential expanded about the $\phi=1$ vacuum is of the form
\beq
V[\phi]=A(1-\phi)^2+B(1-\phi)^{8/3}+\cdots.
\eeq

\section{Discussion}\label{sec:discussion}


We have seen that LSPT can be easily applied to Stokes and anti-Stokes scattering in PT models.  The results for the excitation and de-excitation of the two shape modes in the $\sigma=3$ kink are similar to those found
in the $\sigma=2$ case of the $\phi^4$ double-well potential in Ref.~\cite{mestokes}.  Namely, the probability of each process diverges\footnote{Of course the probability is never more than unity.  As the tree-level probability of order unity, the perturbative expansion becomes invalid.} near its threshold and then decreases monotonically to zero, with backward scattering decreasing faster than forward scattering, where we recall that the threshold for anti-Stokes scattering is $k_0=0$.  When $k_0$ is a fixed amount above the threshold, both excitation and de-excitation are several times more likely in the case of the first shape mode than the second.

More generally, the higher PT scalar models enjoy potentials whose third derivative diverges in the vacuum.  As a result, the three-point coupling that appears in old-fashioned perturbation theory is infinite and so these models have largely been dismissed in the literature.  This is despite the fact that these models, if they are sensible, would provide and excellent toy models for the understanding of quantum solitons, because their normal modes are known in closed form.  So far the Sine-Gordon and $\phi^4$ model have needed to fill this role, but extrapolating from only two models is rather dangerous.

In the present paper we have taken a first step in rehabilitating these models.  We have shown that this same divergent cubic coupling in fact leads to a finite tree-level amplitude for a leading order inelastic scattering processes in which incoming radiation excites or de-excites an internal shape mode.  

At tree level, the divergence canceled.  This does not mean that it can be ignored.  We have found that the divergence arises from a $\phi^{8/3}$ coupling, and three such couplings of course may be combine into an integer power coupling, and so already affects the usual Fock space decomposition of Hamiltonian eigenstates.   

Of course, this interaction will correct Hamiltonian eigenstates states at even lower orders in perturbation theory.  As $\phi^{8/3}$ can not be expanded as a polynomial in the usual creation and annihilation operators, we expect such corrections to lie outside the usual $n$-body Fock space.  But this does not mean that they cannot be constructed.  Indeed, there are many examples of quantum field theories, especially in 1+1 dimensions\footnote{One among many examples is the fermion monopole system that has recently caused controversy \cite{tong,stefano}.}, with such fractional local operators that take one out of the Fock space.  Moreover, in quantum mechanics the spectrum of the $x^{8/3}$ theory is known \cite{znojil} and it may be that the Fock space of the quantum field theory consists of these quantum mechanical bound states, each with their own momentum.  

At leading order, a meson and a reflectionless kink may scatter via three processes, all of which are inelastic.  The first is meson multiplication, in which the final state consists of a kink and two mesons.  The others are Stokes and anti-Stokes scattering, in which the final state consists of a single meson but one of the kink's shape modes is respectively excited or de-excited.  The relevant coupling for all three of these processes is the cubic coupling in the Hamiltonian, which diverges in the vacuum for the $\sigma=3$ theory.  Nonetheless, in the present paper we have found that this divergence does not lead to a divergence in the probability of Stokes or anti-Stokes scattering, as a result of the localization of the shape modes.

It would be interesting to see whether the probability of meson multiplication is also finite.  In this case, the shape mode does not help to localize the interaction, and the absolute value of the integrand in the $V_{kkk}$ interaction diverges exponentially at large distance, although its phase oscillates rapidly.  This process requires a transfer of momentum from the mesons to the kink which suggests that the interaction point must be physically close to the kink, potentially removing the divergence.  Nonetheless, $V_{kkk}$ itself is likely ill-defined.  We suspect that this naive divergence results because the correction to the states due to the $\phi^{8/3}$ interaction is of lower order than the meson multiplication interaction, and so it must be computed first.  Once the leading correction to the state is applied, a meson will no longer evolve in the vacuum at this order and so will only interact in the vicinity of the kink, removing the divergence.  

This is analogous to the fact that quantum mechanics with an $x^2+\lambda x^{8/3}$ potential, one cannot find corrections to the oscillator states using $a^\dag$ and $a$ operators.  Nonetheless, an asymptotic expansion in $\lambda$ is provided by perturbation theory in that case, leading to wave functions which are sums of rational powers of $x$.  This procedure corresponds, in quantum field theory, to perturbation theory directly applied to Schrodinger wave functionals, where such rational powers of $\phi(x)$ are not problematic.  We will leave this approach to meson multiplication in PT models to future work.


\appendix

\section{Decomposition Method for the PT potential} \label{app}

Consider a function $W_n(x)$ called the superpotential, indexed by a number $n$ which corresponding to the $\sigma$ in main text.  Use it to construct the operators
\beq
A_n=\partial_x+W_n(x)\hsp
A_n^\dag=-\partial_x+W_n(x).
\eeq
Then define the Hamiltonian $H_n$ and its dual, conventionally called $H_n^\dag$, as
\beq
H_n\equiv A_n^\dag A_n=-\partial_x^2-W\p_n(x)+W_n^2(x)\hsp H_n^\dag =\partial_x^2+W_n\p(x)+W^2_n(x).
\eeq

Using the identity
\beq
H_nA_n^\dag=A_n^\dag H_n^\dag 
\eeq
one finds that for any eigenmode $\phi$ of $H\p_n$
\beq
H\p_n\phi=E\phi
\eeq
$A^\dag_n\phi$ will be an eigenmode of $H_n$
\beq
H_n A^\dag_n \phi=A^\dag_n H^\dag\phi_n=E A^\dag_n\phi.
\eeq

We are interested in the superpotential
\beq
W_n(x)=an\tanh(ax)
\eeq
so that the Hamiltonian has the PT form 
\beq
H_n=-\partial_x^2-a^2n(n+1)\sech^2(ax)+a^2n^2
\eeq
and the dual is
\beq
H_n^\dag=-\partial_x^2-a^2n(n-1)\sech^2(ax)+a^2n^2=H_{n-1}+a^2(n^2-(n-1)^2).
\eeq
In other words, $H^\dag_n$ and $H_{n-1}$ have the same eigenvectors.

We have learned that if $\phi$ is eigenvector of $H_n$ then it is also an eigenvector of $H^\dag_{n+1}$ and so $A^\dag_n\phi$ is an eigenvector of $H_{n+1}$.  Therefore, given eigenvector $\phi$ of any $H_{n_1}$, we can construct a corresponding eigenvector for $H_{n_2}$.  It is simply $A^\dag_{n_2-1}\cdots A^\dag_{n_1}\phi$.

As $H_n=A^\dag_n A_n$, it is clear that any $\phi_0$ annihilated by $A_n$ will be a zero mode of $H_n$.  The normalizable solutions of
\beq
A_n\phi^{(n)}_0=0
\eeq
are, up to normalization,
\beq
\phi^{(n)}_0=\sech^n(a x).
\eeq
These are the zero modes of the level $n$ PT models.  The shape modes can be constructed using the raising operator on zero modes
\bea
  \phi^{(n)}_m(ax)&=& A_{n}^{\dag}\cdots A_{n-m+1}^{\dag}\phi_{0}^{(n-m)}=\prod_{j=[n]-m+1}^{[n]}\bigg(-\partial_x+aj\tanh{(ax)}\bigg)\phi_{0}^{n-m}\nonumber\\
 \omega_m^2&=&a^2(n^2-(n-m)^2).
\eea
The continuum modes can similarly be produced by acting $A^\dag_n\cdots A^\dag_0$ on the plane waves $e^{ikx}$, which are the normal modes corresponding to the free $n=0$ model.  These have frequency
\beq
\ok{}^2=k^2+a^2 n^2.
\eeq

\section* {Acknowledgement}

\noindent
This work was supported by the Higher Education and Science Committee of the Republic of Armenia (Research Project No. 24RL-1C047).
HY Guo was supported by Sun Yat-sen university international Postdoctoral Exchange Program and also supported by Research Projects Developed by the Lanzhou Theoretical Physics Center/Gansu Provincial Key Laboratory of Theoretical Physics (Research Project: NSFC Grant No. 12247101). HYG and SB are supported  by the INFN special research project
grant ``GAST'' (Gauge and String Theories).

\end{document}